\documentclass[aps,pra,preprint,superscriptaddress,longbibliography]{revtex4-2}

\usepackage{graphicx}
\usepackage{amsmath}
\usepackage{color}

\begin{document}
	
	\title{Harmonic motion modes in parabolic GRIN fibers}

        \author{A. Collado Hern\'andez}
        \email[e-mail: ]{adolfo.colladoh@uanl.edu.mx}
        \affiliation{Facultad de Ciencias F\'isico Matem\'aticas, Universidad Aut\'onoma de Nuevo Le\'on, San Nicol\'as de los Garza, 66455, Mexico}
        
        \author{F. Marroqu\'in}
        \email[e-mail: ]{fco.marroquin.gtz@upp.edu.mx}
        \affiliation{Universidad Polit\'ecnica de Pachuca. Carr. Pachuca-Cd. Sahag\'un Km.20, Ex-Hda. Santa B\'arbara. Zempoala, 43830 Hidalgo, Mexico }
        
        \author{B.~M. Rodr\'iguez-Lara}
        \email[e-mail: ]{bmlara@upp.edu.mx}
        \affiliation{Universidad Polit\'ecnica de Pachuca. Carr. Pachuca-Cd. Sahag\'un Km.20, Ex-Hda. Santa B\'arbara. Zempoala, 43830 Hidalgo, Mexico }
	
	\date{\today}
	
    \begin{abstract}
        We report electromagnetic field modes solving the inhomogeneous Maxwell equations for parabolic gradient index fibers in the low refractive index contrast approximation.
        The first family comprises accelerating fields characterized by an intensity distribution center tracing a circular trajectory transverse to the fiber optical axis. These fields maintain an invariant shape for both their intensity and phase distributions while rotating around their center.
        The second family comprises breathing fields characterized by an intensity distribution center aligned with the fiber optical axis. These fields exhibit intensity distribution scaling along propagation, while their phase swirls and rotates around the optical axis without changing their intensity distribution shape and topological charge.
    \end{abstract}

    \maketitle
    \newpage
	
\section{Introduction}\label{sec1}

Parabolic graded refractive index (GRIN) fibers \cite{Marcuse1973} have a smoothly varying refractive index profile that decreases radially from the fiber axis that provides reduced modal dispersion compared to step-index fibers \cite{Ostermayer1974,Cohen1975}. 
Their normal modes are well-understood with the analytic description of their normal modes, propagation constants, and cladding-mode cutoffs \cite{Marcuse1973,Yamada1977,Garside1980}.
Beyond telecommunications, they find utility in strain and temperature sensing \cite{Tripathi2009,BeltranMejia2014}, as well as
 medical imaging \cite{Bolshtyansky1997, Lorenser2013} to mention a few examples.

In the ideal scenario of infinite parabolic GRIN media, the propagation of Hermite- and Laguerre- \cite{Tien1965, Newstein1987}, as well as Ince-Gauss \cite{GutierrezVega2005} modes have been studied, leading to the discovery of hypergeometric breathing  \cite{Kotlyar2013} and pendulum-type \cite{Jia2023} modes.  

Here, we focus on realistic  low-contrast parabolic GRIN fibers to propose electromagnetic field modes that solve Maxwell equations for inhomogeneous media. 
Employing the weak guidance approximation, we build analytic propagation invariant modes with constant radial and azimuthal numbers, find their propagation constants and cut-off relations that are in good agreement with numerical results from finite element method (FEM).
We identify symmetries related to the conservation of circular and azimuthal numbers to construct two families of harmonic motion modes using analogies to coherent states in quantum optics. 
One family are accelerating light fields whose intensity distribution center follows a circular trajectory in the plane transverse to propagation. 
The other are breathing light fields whose intensity distribution width varies with propagation.

\section{Propagation invariant modes}\label{sec2}

\begin{figure}
	\centering
	\includegraphics[scale=1]{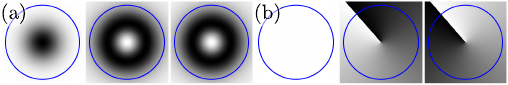}
	\caption{ (a) Intensity $\vert \Psi_{p,\ell}(\rho, \varphi) \vert^2$, and (b) phase $ \mathrm{arg}\left[ \Psi_{p,\ell}(\rho, \varphi)\right]$ distributions for the three modes, $p=0, \ell = 0, \pm 1$, supported by a parabolic GRIN fiber with cladding refractive index $n_{\mathrm{cl}} = 1.444$, refractive index contrast $\Delta n = 0.05$, and core radius $a = 1.058 \, \mu \mathrm{m}$ shown in blue, for  free space wavelength $\lambda_{0} = 632 \, \mathrm{nm}$ provided by FEM.
    } \label{fig:Fig1}
\end{figure}

Consider a parabolic GRIN fiber characterized by a  refractive index \cite{Marcuse1973}, 
\begin{align}
    n(\rho) = n_{\mathrm{cl}} + \Delta n f(\rho), \qquad f(\rho) = \begin{cases}
    1 - \frac{\rho^2}{a^2}, & 0 \le \rho  \le a, \\
    0, &  \rho  > a,
    \end{cases}
\end{align}
where the core of radius $a$ has maximum refractive index $n_{\mathrm{max}} = n(0) = n_{\mathrm{cl}} + \Delta n$ at its optical axis.
The cladding has constant refractive index $n_{\mathrm{cl}}$, resulting in a refractive index contrast $\Delta n = n_{\mathrm{max}} - n_{\mathrm{cl}}$.
For low refractive index contrast, $\Delta n / n_{\mathrm{cl}} \ll 1$, we use the weak guidance approximation \cite{Snyder1983} to propose propagation invariant transverse electromagnetic fields, 
\begin{align}
    \begin{aligned}
        \mathbf{E} \approx& ~\hat{\mathbf{e}} \vert E \vert  e^{i \theta} \Psi(\rho, \varphi) e^{ i \left( \beta z - \omega t \right)} , \\
        \mathbf{B} \approx&~ \left( \hat{\mathbf{z}}\times \hat{\mathbf{e}} \right) \frac{n(\rho)}{c}  \vert E \vert  e^{i \theta} \Psi(\rho, \varphi) e^{ i \left( \beta z - \omega t \right)},
    \end{aligned}
\end{align}
that solve Maxwell equations for an inhomogeneous medium \cite{Chew1995}.
The fields are given in terms of the unit polarization vector $\hat{\textbf{e}}$.
real field amplitude $\vert E \vert $ with phase $\theta$, speed of light in vacuum $c$ with field frequency $\omega$, propagation constant $\beta$, low contrast approximated refractive index, 
\begin{align}
    n^{2}(\rho) \approx n_{\mathrm{cl}}^{2} + 2 \Delta n f(\rho),
\end{align}
and scalar wave function,
\begin{align}
\Psi_{p, \ell}(\rho,\varphi) = A \left\{ \begin{aligned}
& \psi_{p,\ell}(\rho, \varphi)  & 0\leq \rho \leq a, \\
& \frac{\psi_{a,\ell}(a, \varphi) }{K_{ \vert \ell \vert } \left( \alpha  a \right) } K_{ \vert \ell \vert } \left( \alpha  \rho \right) e^{i \ell \varphi}  & \rho > a ,
\end{aligned} \right.
\end{align}
solving the Sturm-Liouville eigenvalue problem, 
\begin{align}
    \left[\nabla^2_t + 2 k_0^2 n_{\mathrm{cl}} \Delta nf(x_1,x_2) - \beta^{2} + k_{0}^{2} n_{\mathrm{cl}}^{2} \right]\Psi(x_1,x_2)=0,
\end{align}
in terms of Laguerre-Gauss modes, 
\begin{align}
    \psi_{p, \ell}(\rho, \varphi) 
    = (-1)^{p} \sqrt{\frac{ p!}{\pi  (p+\vert \ell \vert)!}} \, \left(\frac{\rho}{\sigma}\right)^{\vert \ell \vert} e^{ - \frac{\rho^{2}}{2 \sigma^{2}} } \mathrm{L}_{p}^{\vert \ell \vert} \left( \frac{\rho^{2}}{\sigma^{2}} \right) e^{i \ell \varphi},
\end{align}
with generalized Laguerre polynomials $L_{n}^{(\gamma)}(x)$, constant Gaussian waist, 
\begin{align}
    \sigma^{2} = \frac{a}{k_{0}} \sqrt{\frac{1}{2 n_{\mathrm{cl}} \Delta n}},
\end{align}
and propagation constant, 
\begin{align}
    \beta^{2}(p, \ell) = k_{0}^{2} n_{\mathrm{cl}} \left( n_{\mathrm{cl}} + 2 \Delta n \right) - \frac{2 }{\sigma^{2}} \left( 2p + \vert \ell \vert + 1 \right),
\end{align}
with radial and azimuthal number values $p = 0, 1, 2, \ldots$ and $\ell = -p, -p+1, \ldots, p-1, p$, in that order.
The solution in the cladding is given in terms of modified Bessel functions of the second kind $K_{\gamma}(x)$ with auxiliary propagation constant, 
\begin{align}
    \alpha^{2} = 2 k_{0}^{2} n_{\mathrm{cl}} \Delta n - \frac{2}{\sigma^{2}} (2 p + \vert \ell \vert + 1),
\end{align}
vacuum wave number $k_{0} = 2 \pi / \lambda_{0}$, and vacuum wavelength $\lambda_{0}$.
These propagation constants provide the cut-off relation,  
\begin{align}
     k_{0} a \ge  \sqrt{\frac{2}{n_{\mathrm{cl}} \Delta n}} \left( 2p + \vert \ell \vert + 1 \right) ,
\end{align}
for the number of modes supported by a particular fiber .
We choose the normalization parameter $A$ to provide, 
\begin{align}
    \int d^{2}r ~ \Psi_{p^{\prime},\ell^{\prime}}^{\ast}( \rho, \varphi ) \Psi_{p,\ell}( \rho, \varphi ) = \delta_{p, p^{\prime}} \delta_{\ell, \ell^{\prime}},
\end{align}
orthonormal modes.
Figure \ref{fig:Fig1} shows the three modes supported by a parabolic GRIN fiber with cladding refractive index $n_{\mathrm{cl}} = 1.444$, refractive index contrast $\Delta n = 0.05$, and core radius $a = 1.058 \, \mu \mathrm{m}$ for free space wavelength $\lambda_{0} = 632 \, \mathrm{nm}$ calculated using FEM simulation that are in good agreement with our analytic results, $\beta(0,0) = 1.460 \times 10^{7} \, \mathrm{rad} \cdot \mathrm{m}^{-1}$ [$\beta_{FEM}(0,0) = 1.461 \times 10^{7} \, \mathrm{rad} \cdot \mathrm{m}^{-1}$] and $\beta(0,\pm 1) = 1.436 \times 10^{7} \, \mathrm{rad} \cdot \mathrm{m}^{-1}$ [$\beta_{FEM}(0,\pm 1) = 1.439 \times 10^{7} \, \mathrm{rad} \cdot \mathrm{m}^{-1}$], up to a rotation of the reference frame.

\section{Harmonic motion modes}\label{sec3}

The normal modes in the core are the optical equivalent of the homogeneous, two-dimensional harmonic oscillator Laguerre-Gauss modes.
Thus, we build  electromagnetic field modes,
\begin{align}
    \begin{aligned}
        \mathbf{E} \approx& ~\hat{\mathbf{e}} \vert E \vert  e^{i \theta} \Phi_{\pm}(k, \delta; \rho, \varphi,z) e^{ - i \omega t} , \\
        \mathbf{B} \approx&~ \left( \hat{\mathbf{z}}\times \hat{\mathbf{e}} \right) \frac{n(\rho)}{c}  \vert E \vert  e^{i \theta} \Phi_{\pm}(k, \delta; \rho, \varphi,z) e^{ -i  \omega t},
    \end{aligned}
\end{align}
with polarization that remains invariant to propagation and in terms of scalar wave functions, 
\begin{align}
    \begin{aligned}
        \Phi_{\pm}(k, \delta; \rho, \varphi, z) =&~ A \sum_{m = 0}^{\infty} \sqrt{P(k,m,\delta)}~  \Psi_{p, \pm \ell}(\rho, \varphi)  
          e^{ i  \beta(p, \pm \ell) z}, \\
    \end{aligned}
\end{align}
that are the superposition of propagation invariant modes weighted by a probability distribution where radial and azimuthal numbers may depend on the summation index $m$; that is, $p \equiv p(k,m)$ and $\ell \equiv \ell(k,m)$.
The probability distribution dictates the number of involved modes, while our cut-off relation provides us with the fiber radius size needed to support them.
In addition,  we approximate our analytic propagation constant,
\begin{align}
    \begin{aligned}
    \beta(p,\ell) \approx&~ \beta_{0} - \beta_{1}(p,\ell), \\
    \beta_{0} =&~ k_{0} \sqrt{n_{\mathrm{cl}} \left( n_{\mathrm{cl}} + 2 \Delta n \right)}, \\
    \beta_{1}(p,\ell) =&~\frac{1}{a} \sqrt{ \frac{2  \Delta n }{n_{cl}+2\Delta n}}~ \left( 2 p + \vert \ell \vert + 1\right) ,
    \end{aligned}
\end{align}
in order to simplify our analysis. 
This approximation highlights the core radius maintains accuracy by balancing the contribution of higher order field modes. 

\section{Harmonic motion modes type I}\label{sec4}

In the two-dimensional quantum harmonic oscillator, circular numbers, $n_{+}$ and $n_{-}$, define the radial and azimuthal numbers, $p = \mathrm{min}(n_{+},n_{-})$ and $\ell = n_{+} - n_{-}$.
Subspaces with constant circular numbers $n_{\pm}$ exhibit an underlying Heisenberg-Weyl symmetry \cite{RodriguezMorales2023}.
Following the Glauber-Suddarshan coherent states  \cite{Glauber1963, Sudarshan1963} recipe yields a Poisson probability distribution, radial, and azimuthal numbers, 
\begin{align}
    \begin{aligned}
        P(m, \delta) =&~ e^{-\vert \delta \vert^{2}} \frac{\vert \delta \vert^{m}}{m!}, \\
        p(k,m) =&~\mathrm{min}(k,m), \\
        \ell(k,m) =&~ \pm (k-m),
    \end{aligned}
\end{align}
where the complex parameter $\delta$ amplitude and phase provide the intensity distribution center radial and angular displacement, respectively, in the plane transverse to propagation.
These states have topological charge equal to $\ell = \pm k$. 

\begin{figure}
	\centering
	\includegraphics[scale=1]{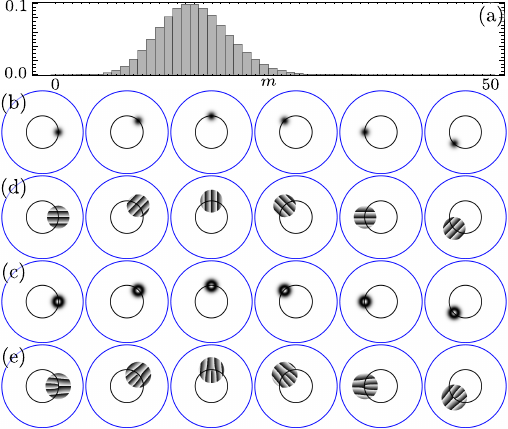}
	\caption{ (a) Poisson probability distribution $P(m,\delta)$ for a parameter $\delta = 4$. 
    (b) [(c)] Intensity and (d) [(e)] phase distribution for our harmonic motion mode with topological charge $k=0$ [$k=1$] at propagation distances $\beta_{1}(0,0) z =  0, \pi/4, \pi/2, 3 \pi / 4, \pi, 5 \pi/4$the .
    Core radius $a = 28.058 \, \mu\mathrm{m}$ shown in blue and radial displacement $\rho_{0}(z) = 11.072 \, \mu \mathrm{m}$ [$\rho_{0}(z) = 11.246 \, \mu \mathrm{m}$] in black.
    } \label{fig:Fig2}
\end{figure}

It is simpler to generate insight using mode with $k=0$, 
\begin{align}
    \Phi_{\pm}(k, \delta; \rho, \varphi, z) = &~ \frac{A}{\sqrt{\pi}} e^{-\frac{\vert \tilde{\delta}(z) \vert^{2}}{2}}  e^{- \frac{\rho}{\sigma} \tilde{\delta}(z)  } e^{- \frac{\rho^{2}}{2 \sigma^{2}}} , \quad \rho \le a,
\end{align}
with modified parameter, 
\begin{align}
    \tilde{\delta}(z) = \delta e^{  i \beta_{1}(0,0) z},
\end{align}
that takes the form of a displaced Gaussian scalar wave within the core.
The intensity distribution center locates at the radial distance, 
\begin{align}
    \rho_{0} = \int_{0}^{\infty} d\rho \int_{0}^{2 \pi} d\varphi~  \rho^{2} \vert \Phi_{\pm}(k, \delta; \rho, \varphi, z) \vert^{2},
\end{align}
in terms of the modified Bessel function of the first kind $I_{\gamma}(x)$.
It is straightforward to see that the intensity distribution center follows an helical trajectory with constant radius $\rho_{0}(z) = \rho_{0}$, and angular position $\varphi_{0}(z) = \mathrm{arg} (\delta) - \beta(0,0) z$ proportional to the propagation distance.
Its intensity and phase distributions in the plane transverse to propagation rotate around the point defined by these functions.
Figure \ref{fig:Fig2}(a) shows Poisson probability distribution for a parameter $\delta = 4$.
It is enough to consider the first fifty modes $m \in [0,50]$ to construct our states.
Considering a parabolic GRIN fiber with cladding refractive index $n_{\mathrm{cl}} = 1.444$ and refractive index contrast $\Delta n = 0.05$ for free space wavelength $\lambda_{0} = 632 ~\mathrm{nm}$, yields a minimum core radius $a = 28.058 \mu\mathrm{m}$ for a fiber supporting topological charges $k=0,1$.
The maximum absolute error between the analytic and approximated propagation constant for these parameters and modes  $\beta = \vert \beta(k,50) - \beta_0 + \beta_{1}(k,50) \vert / \beta(k,50) \approx 5 \times 10^{-4}$ is negligible.
Figure \ref{fig:Fig2}(b) [Fig. \ref{fig:Fig2}(d)] shows the intensity and Fig. \ref{fig:Fig2}(c) [Fig. \ref{fig:Fig2}(e)] the phase distributions for our harmonic motion mode with topological charge $\ell = k = 0$ [$\ell = k = 1$] at small propagation distances. 

\section{Harmonic motion modes type IIa}\label{sec5}

\begin{figure}
	\centering
	\includegraphics[scale=1]{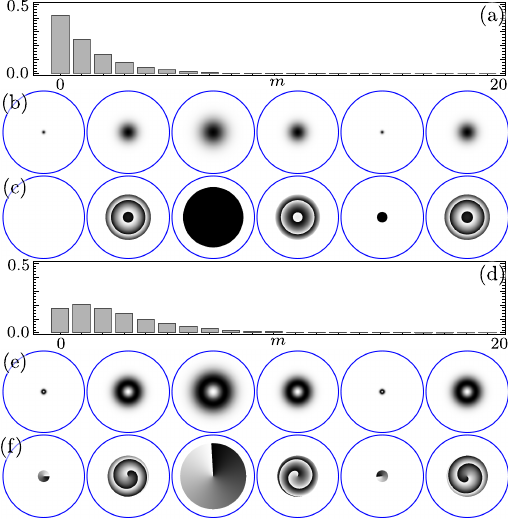}
	\caption{ (a)[d] Probability distribution $P(m,\delta)$ for a parameter $\delta = 2$. 
    (b) [(e)] Intensity and (d) [(f)] phase distribution for our harmonic motion mode with topological charge $k=1/2$ [$k=1$] at propagation distances $\beta_{1}(0,0) z = 0, \pi/4, \pi/2, 3 \pi / 4, \pi, 5 \pi/4$.
    Core radius $a = 22.235 \, \mu\mathrm{m}$ shown in blue.
    } \label{fig:Fig3}
\end{figure}

Subspaces with constant azimuthal number $\ell$ exhibit an underlying symmetry provided by the $su(1,1)$ algebra \cite{RodriguezMorales2023}. 
Following the Gilmore-Perelomov coherent states \cite{Gilmore1972,Perelomov1972} recipe yields a probability distribution, radial and azimuthal numbers, 
\begin{align}
    \begin{aligned}
        P \left([ m, \delta \right) =&~ \frac{(2k + m -1)!}{m! (2k-1)!} 
         \mathrm{sech}^{4k} \frac{\vert \delta \vert}{2} \left\vert -\frac{\delta}{\vert \delta \vert} 
         \mathrm{tanh} \frac{\vert \delta \vert}{2} \right\vert^{2m}  , \\
        p(k,m) =&~ m, \\
        \ell(k,m) =&~ \pm(2k-1),
    \end{aligned}
\end{align}
with Bargmann parameter $k= 1/2, 1, 3/2, \ldots$ 
These states take the form of a Gaussian envelope times the radial coordinate raised to the power of the azimuthal number,
\begin{align}
    \begin{aligned}
        \Phi_{\pm}(k, \delta; \rho, \varphi, z) =&~ \frac{A e^{i \left[ \beta_{0} z - i 2 k \beta_{1}(0,0) z \right]}}{\sqrt{\pi (2k-1)!} w(\delta, z)^{ 2 k}} \left( \frac{\rho}{\sigma} \right)^{2k-1} \times \\ 
        &~ \times e^{-\frac{\rho^2}{2 \sigma^{2} w^{2}}} e^{ i \phi_{\pm}(\delta,z) }, \quad \rho \le a,
    \end{aligned}
\end{align}
within the core.
The Gaussian envelope waist and overall phase,
\begin{align}
    \begin{aligned}
    w^{2}(\delta,z) =&~ \cosh \vert \delta \vert - \cos \{ \arg[\tilde{\delta}(z)] \} \sinh \vert \delta \vert, \\
    \phi_{\pm}(\delta,z) =&~  \pm (2k -1) \varphi -  \sin \{ \arg[\tilde{\delta}(z)] \} \sinh \vert \delta \vert \frac{\rho^{2}}{w^2 \sigma^2},
    \end{aligned}
\end{align}
in that order, in terms of the modified parameter, 
and modified parameter, 
\begin{align}
    \tilde{\delta}(z) = \delta e^{- i 2 \beta_{1}(0,0) z}
\end{align}
show that the intensity distribution width varies with propagation, expanding and contracting, while the pase exhibits a topological charge $\ell = \pm (2k -1)$. 
The phase distribution is a function of the radial coordinate and propagation distance, transitioning from a standard to a swirl vortex with propagation, all while rotating around the fiber optical axis.
Figure \ref{fig:Fig3}(a) [Fig. \ref{fig:Fig3}(d)] shows the intensity distribution for a parameter $\delta = 2$.
It is enough to consider the first twenty modes $m \in [0,20]$ to construct our states.
Using the same fiber yields a minimum core radius $a = 22.235 \mu\mathrm{m}$ for a fiber supporting topological charges $k=0,1$.
The maximum absolute error between the analytic and approximated propagation constant for these parameters and modes  $\beta = \vert \beta(k,50) - \beta_0 + \beta_{1}(k,50) \vert / \beta(k,50) \approx 4 \times 10^{-3}$ is negligible.
Figure \ref{fig:Fig3}(b) [Fig. \ref{fig:Fig3}(e)] shows the intensity and Fig. \ref{fig:Fig3}(c) [Fig. \ref{fig:Fig3}(f)] the phase distributions for our harmonic motion mode with topological charge $\ell = 2k -1 =0$ [$\ell = 2 k -1 = 1$] at small propagation distances.

\section{Harmonic motion modes type IIb}\label{sec6}

\begin{figure}
	\centering
	\includegraphics[scale=1]{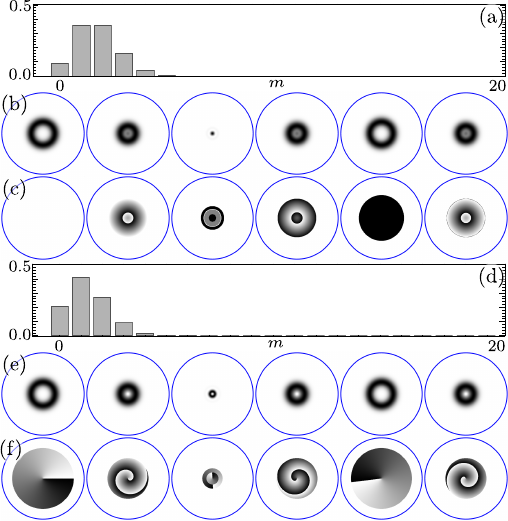}
	\caption{ Same as Fig. \ref{fig:Fig3} for Bessel-Gauss mode.
    } \label{fig:Fig4}
\end{figure}

In the same subspaces, following the Barut-Girardello coherent state \cite{Barut1971} recipe yields a probability distribution, radial and azimuthal numbers, 
\begin{align}
    \begin{aligned}
        P \left(k,  m,  \delta \right) =&~ \frac{\vert \delta \vert^{(2k + 2m - 1)}}{(2k-1)! \, m! \, \mathrm{I}_{2k-1}(2 \vert \delta \vert)}  , \\
        p(k,m) =&~ m, \\
        \ell(k,m) =&~ \pm(2k-1),
    \end{aligned}
\end{align}
with Bargmann parameter $k=1/2, 1, 3/2, \ldots$
These states take the form of a Bessel-Gauss scalar wave function, 
\begin{align}
    \begin{aligned}
        \Phi_{\pm}(k, \delta; \rho, \varphi, z) =&~ \frac{A (-1)^{-k + \frac{1}{2}} e^{-\tilde{\delta}(z)}}{\sqrt{ \pi \, \mathrm{I}_{2k-1}(2 \vert \delta \vert)}} e^{i \left[ \beta_{0} -  i 2 k \beta_{1}(0,0) \right] z} 
        e^{-\frac{\rho^{2}}{2 \sigma^{2}}} \times \\
       &~ \times \mathrm{J}_{2k - 1}\left[  \, \frac{2 \sqrt{- \tilde{\delta}(z)}}{ \sigma} \rho \right] e^{ \pm i (2k -1) \varphi}, \, \rho \le a,
    \end{aligned}
\end{align}
within the core in terms of the Bessel function of the first kind $\mathrm{J}_{\gamma}(x)$ with complex argument, 
\begin{align}
    \tilde{\delta}(z) = \delta e^{- i 2 \beta_{1}(0,0) z} . 
\end{align}
It is not straightforward to see, but the intensity distribution width varies with propagation, expanding and contracting, while the phase shows a topological charge $\ell = \pm (2k -1)$.
The phase distribution of the Bessel function with complex argument depends on the radial coordinate and propagation distance, transitioning from a standard to a swirl vortex with propagation, all while rotating around the fiber optical axis.
Figure \ref{fig:Fig4}(a) [Fig. \ref{fig:Fig4}(d)] shows the intensity distribution using the same fiber as before.
Figure \ref{fig:Fig4}(b) [Fig. \ref{fig:Fig4}(e)] shows the intensity and Fig. \ref{fig:Fig4}(c) [Fig. \ref{fig:Fig4}(f)] the phase distributions for our harmonic motion mode with topological charge $\ell = 2k -1 =0$ [$\ell = 2 k -1 = 1$] at small propagation distances.

\section{Conclusion}\label{sec7}

In summary, we showed that low-contrast parabolic GRIN fibers support field modes akin to coherent states for two intrinsic symmetries involving circular and radial number conservation.
We introduced two families of harmonic motion field modes.
One exhibiting an intensity and phase distribution center following helical trajectories of constant radius along propagation while remaining invariant. 
The other shows varying intensity distribution width through propagation, yet retains its shape and topological charge while its phase distribution transitions from standard to swirl vortex. 
Our analysis shows that  larger the core radii yield a closer match between numerical FEM and analytic propagation constants, enabling our harmonic motion fields to propagate over longer distances without distortion; for example, a fiber core radius $a = 1 \, \mathrm{mm}$ allows propagation distances spanning a few meters obviating the crosstalk induced by deformations in the fiber.
    
\section*{Funding}
 A. C. H. doctoral studies funded by CONAHCYT grant 841625.   
\section*{Acknowledgments}	
B.~M.~R.-L. thanks Dilia Aguirre Olivas, Gabriel Mellado-Villase\~nor, Benjamin Raziel Jaramillo Avila, Zulema Gress Mendoza, and Benjamin Perez-Garcia for fruitful discussions

	
\section*{References}

%

\end{document}